\documentstyle[iopconf,axodraw,psfig]{article}
\newcommand{\DR}{\mbox{{\footnotesize{$\overline{{\rm DR}}$}}}}

\def\gsim{\kern.4em\raise.3ex
\hbox{$>$\kern-.75em\lower1ex\hbox{$\sim$}}\kern.4em}
\def\lsim{\kern.4em\raise.3ex
\hbox{$<$\kern-.75em\lower1ex\hbox{$\sim$}}\kern.4em}
\def\psla{p\kern-.45em/}
\def\ssla{S\kern-.45em/}
\begin{document}
\title{Toward {\it Loop Level} Study of Supersymmetric Model at 
Future $e^+e^-$ Linear Collider}
\author{Mihoko M. Nojiri \footnote{Address after Oct.1st. 
Yukawa Institute, Kyoto University, Kyoto, Japan.  
E-mail:nojirim@theory.kek.jp.
This work is supported in part by Grant in aid for 
Science and Culture of Japan(07640428, 09246232)}} 

\affil{KEK Theory Group, Oho 1-1, Tsukuba, Ibaraki, 305,Japan}

\beginabstract
In the
case with a large splitting between the squark and slepton masses, 
the supersymmetric identities which enforce the equality of
the gauge and gaugino couplings are violated in the effective 
theory below the squark mass threshold.  
We compute the full one-loop (s)quark corrected slepton
production cross-sections. 
We find that the one-loop corrected
slepton production cross-sections can depend on the squark mass
strongly, up to 
$9\% \times\log_{10}(M_{\tilde{Q}}/m_{\tilde{\ell}} )$.  
We investigate the squark
mass sensitivity of the slepton cross-section measurements at a future
linear collider.  For sneutrino production accessible at 
$\sqrt{s}=500$ GeV there can be sensitivity to squark 
masses at or larger than 1
TeV.
\endabstract

\section{Introduction}
Supersymmetry is an attractive possibility beyond the standard model.
Because of the relations supersymmetry imposes among the dimensionless
couplings, the quadratic divergences in the Higgs sector are cut-off
by the superpartner mass scale.  The cancellations stabilize the
hierarchy between the Planck scale and the weak scale.  The minimal
supersymmetric standard model (MSSM) is consistent with gauge coupling
unification suggested by grand unified theories.  Also, it is
interesting that one of the hallmarks of supersymmetry, a light Higgs
boson ($m_h\lsim130$ GeV), is favored by global fits to precision
electroweak data

In this contribution we examine the prospect of testing supersymmetry via a
precise measurement of the lepton-slepton-gaugino vertex. The linear
collider provides a suitably clean experimental environment
\cite{JLC1,TSUKA,FPMT,BMT,NFT}.  Among
the relations which account for the cancellations of quadratic
divergences, supersymmetry relates the lepton-slepton-gaugino
coupling to the usual gauge coupling.

Although bare (or \DR) couplings enjoy the relations imposed by
supersymmetry, the effective gauge and gaugino couplings are not equal
because supersymmetry is broken. In particular, all non-singlet
nondegenerate supermultiplets such as the quark-squark supermultiplets
contribute to the splitting.  Hence, measurements of the type we consider here
not only provide for detailed tests of supersymmetry, but can also
elucidate important features of the scale and pattern of supersymmetry
breaking \cite{CFPA,CFPB,RKS,PT,NPY}.

For example, the (s)quark contribution to the splitting of the U(1)
and SU(2) gaugino/gauge (s)lepton couplings grows logarithmically with
the squark mass, as
\begin{equation}
\label{ln msq}
{\delta {g_Y} \over {g_Y}} \simeq {11g_Y^2\over48\pi^2}\ln\left(
{M_{\tilde Q}\over m_{\tilde \ell}}\right)\ ,
\qquad\qquad {\delta g_2\over g_2} \simeq
{3g_2^2\over16\pi^2}\ln\left(M_{\tilde Q}\over m_{\tilde\ell}\right)\ .
\end{equation}
This correction is obtained by evolving the couplings according to the
renormalization group equations (RGE's) of the effective theory
\cite{C} below the squark mass threshold.  When $M_{\tilde
Q}/m_{\tilde\ell}\simeq 10$ the correction to the SU(2) (U(1))
coupling is about 2\% (0.7\%). This gives rise to an enhancement of
the $t$-channel slepton or gaugino production cross-section of about
8\% (2.8\%).  If large statistics are available and systematic errors
can be controlled, we can (assuming the MSSM) constrain the squark
mass scale by this measurement. 

We restrict our attention to the measurement of the first generation
lepton-slepton-gaugino coupling at an $e^-e^+$ linear collider. Much
study has been undertaken to determine how accurately we can expect to
measure these couplings\cite{FPMT,NFT, CFPB, NPY}.  
In Ref.\cite{NPY}, we perform a full one-loop calculation 
of the slepton production
cross-section within the MSSM. 
We include only (s)quark loops in the calculation, because the
correction is enhanced by a color factor and the number of
generations. The remaining corrections are small, and if we did
include them we expect our conclusions would not change.

In this contribution,  we only discuss our calculation 
only briefly in section 2.
We point out that, to a good approximation, the one-loop $t$-channel
amplitudes can be rewritten in the same form as the tree-level
amplitudes, with the replacement of the tree-level parameters with
renormalized effective parameters. Hence we introduce the effective
coupling, the effective masses, and the effective mixing matrix.  
In section 3 we discuss our 
numerical results, and show how well we can measure the squark loop
correction to the coupling, and thereby constrain the squark mass,
assuming both slepton and chargino production are possible.  We show
the statistical significance of the results by combining our knowledge
of the superpartner masses and cross-sections.    The
uncertainty in the slepton mass measurement is quite important in this
analysis.   In the
last section, section 4, we give our conclusions.
\section{calculation}
In this section we discuss the calculation of the cross-section of
$e^+e^-\rightarrow\tilde\ell_i\tilde\ell_j^*$, where
$\tilde\ell_i=(\tilde e_L^-,\tilde e_R^-,\tilde\nu_e)$, including
one-loop (s)quark corrections. The full result is explicitly given in
Ref.\cite{NPY}; here we restrict ourselves to outline the general features
of the calculation.

The tree level slepton productions proceed through  s-channel 
exchange of $Z$ and $\gamma$, and t-channel exchange of neutralinos
or charginos ($\tilde{\chi}_i$).
To evaluate the one-loop amplitude, we treat all the parameters
appearing in this tree-level expression as running \DR~  quantities, and
add the contributions from the one-loop diagrams (see Fig.~1).  Note
that the (s)quark loop corrections do not give rise to external
wave-function renormalization.

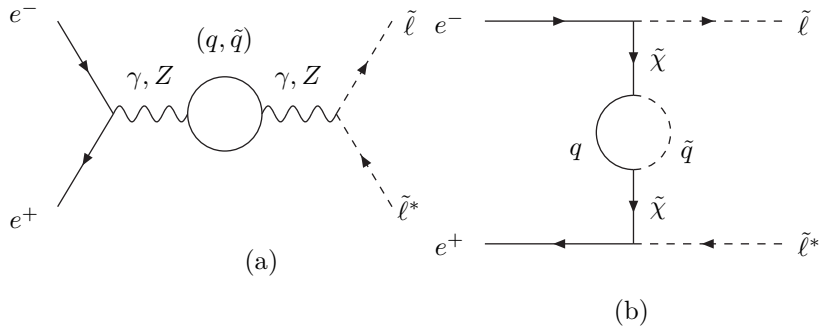
\begin{figure}
\begin{picture}(308,133)(0,7)
% s-channel loop
\ArrowLine(7, 119)(28, 84) 
\ArrowLine(28, 84)(7,49)  
\DashArrowLine(133,49)(112,84){3} 
\DashArrowLine(112,84)(133,119){3} 
\Photon(28,84)(56,84){3}{3} 
\BCirc(70,84){14} 
\Photon(84,84)(112,84){3}{3} 
\Text(0,119)[rb]{$e^-$}
\Text(0,49)[rt]{$e^+$}
\Text(140,119)[]{$\tilde{\ell}$}
\Text(140,49)[]{$\tilde{\ell}^*$}
\Text(70,112)[]{$(q,\tilde{q})$}
\Text(42,98)[]{$\gamma,Z$}
\Text(98,98)[]{$\gamma,Z$}
\Text(84,28)[]{(a)}
% t-channel
\ArrowLine(168,119)(224,119) % (0 190)
\ArrowLine(224,119)(224,91)
\CArc(224,77)(14,90,270)
\DashCArc(224,77)(14,270,90){3}
\ArrowLine(224,63)(224,35)
\ArrowLine(224,35)(168,35)
\DashArrowLine(224,119)(280,119){3}
\DashArrowLine(280,35)(224,35){3}
\Text(161,119)[r]{$e^-$}
\Text(161,35)[r]{$e^+$}
\Text(287,119)[l]{$\tilde{\ell}$}
\Text(287,35)[l]{$\tilde{\ell}^*$}
\Text(203,70)[]{$q$}
\Text(245,70)[]{$\tilde{q}$}
\Text(231,49)[l]{$\tilde{\chi}$}
\Text(231,105)[l]{$\tilde{\chi}$}
\Text(224,14)[t]{(b)}
\end{picture}
\caption{Feynman graphs of the one-loop quarks-squark corrections to 
the processes $e^-e^+\rightarrow\tilde{\ell}\tilde{\ell}^*$, 
for (a) $s$-channel and (b) $t$-channel amplitudes. }
\end{figure}

To avoid a complexities of several gauge interactions, 
let us consider $e^+e^-\rightarrow \tilde{e}^+_R\tilde{e}^-_R$ 
production. If the $\sqrt{s}\gg m_Z$, the process approximately 
proceeds through s-channel exchange of $B$ boson which 
couples to hypercharge,  and 
t-channel exchange of $\tilde{B}$, superpartner of $B$ boson.
The s-channel amplitude below squark mass threshold is very 
well approximated by effective coupling $g_Y^{\rm eff}(Q=\sqrt{s})$. 
$g_Y^{\DR}(Q)$ in MSSM is related to $g_Y^{\rm eff}$ 
as $\alpha_Y^{\rm eff}=$ 
$\alpha^{\DR}(1+\Sigma^q(Q)+\Sigma^{\tilde{q}}(Q))$
where $\Sigma^q(Q)$ and $\Sigma^{\tilde{q}}(Q)$ is 
the gauge two point function from quark and squark loops 
respectively and Q is renormalization scale.
The $g^{\DR}_Y(Q)$ is equivalent 
to the t-channel coupling $g_{\tilde{B}}^{\DR}(Q)$, but 
we do not directly measure the coupling. The 
t-channel amplitude is the sum of tree level contribution and 
1-loop contribution shown in Fig1.(b). The difference 
$\ssla_Y\equiv (g^{\rm eff}_{\tilde{B}}-g^{\rm eff}_Y)/g^{\rm eff}_Y$
$\equiv (\Sigma^{q\tilde{q}}-\Sigma^{q}-\Sigma^{\tilde{q}})(Q)/2$ 
is the correction to the SUSY relation(Fig.2).
 The leading logarithms of the corrections
$\ssla$ at $Q=m_{\tilde\ell}$ are exactly those of
Eq.~(1), showing that the RGE approach of Ref.~\cite{C} gives the proper
results. 
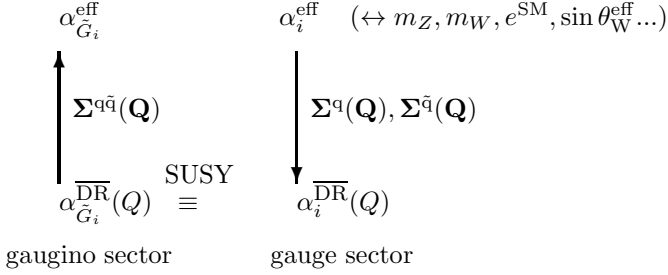
\begin{figure}
\begin{picture}(250,100)(0,-10)  
\thicklines
\put(130, 80){\vector(0,-1){50}}
\put(135, 55){${\bf \Sigma^{\rm q}(Q),\Sigma^{\rm \tilde{q}}(Q)} $ }
\put(45,55){${\bf\Sigma^{\rm q\tilde{q}}(Q)}$}
\put(40,30){\vector(0,1){50}}
\put(120,90){ ${\alpha_{i}^{\rm eff}}$ }
\put(40,90){${\alpha_{\tilde{G}_i}^{\rm eff}}$ }
\put(130,20){${\alpha_i^{\DR}}(Q)$}
\put(40,20){${\alpha_{\tilde{G}_i}^{\DR}(Q)}$}
\put(85,20){${\equiv}$}
\thinlines
%\put(145,90){\line(1,0){20}}
\put(150,90){$(\leftrightarrow m_Z,m_W,e^{\rm SM},
\sin\theta^{\rm eff}_{\rm W}...$)}
\put(120, 0){gauge sector}
\put(80,30) {SUSY}
\put(20, 0){gaugino sector}
\end{picture}
\caption{Schematic figure of the correction to 
SUSY relation}
\end{figure}

Although both s-channel and t-channel diagrams 
receive one loop corrections from squark and quark loops, 
only t-channel amplitude receive physically interesting
correction.  Notice in our approximation, s-channel amplitude 
receives oblique correction of the gauge two point 
function only, which is common to the $e^+e^-$ collision
at Z pole.  The measurement at Z pole fixes the s-channel amplitude 
of slepton production, thanks to the gauge symmetry.

In  Ref\cite{NPY}, we also how the one-loop
corrected $t$-channel
amplitude can be well approximated by a tree-level form. 
For $e^+e^-\rightarrow \tilde{e}_R\tilde{e}_R$, we find
\begin{equation}
{\cal M}^t_{RR} = 2\bar
v\psla\biggl[-\sum_{i=1}^4\frac{\bar{g}^2_{e\tilde{e}_R\tilde{B}}
\overline{N}_{i1}^*\overline{N}_{i1}(p^2) }
{p^2-\overline{m}_i^2(p^2)}P_R\biggr]u\ , \label{eff amp}
\end{equation}
where $p$ is t-channel momentum and 
$\bar{g}_{e\tilde{e}_R \tilde{B}}(p^2)$ 
is the effective bino coupling defined as 
\begin{equation}
\bar{g}_{e\tilde{e}_R \tilde{B}}(p^2)=\hat g_Y(Q)
\biggl(1-\frac{1}{2}\tilde{\Sigma}^L_{11}(Q, p^2)\biggr)\ ,
\end{equation}
and $\tilde{\Sigma}^L_{11}(p^2)$ is the bino-bino component of the
neutralino two-point function, and $\hat g_Y(Q)$ is \DR\  
coupling.  The $\overline{N}_{ij}$ and $\overline{m}_i$ 
are the effective neutralino
mixing matrix and neutralino masses obtained by diagonalizing the
effective neutralino mass matrix $\overline{Y_{ij}}$. $\bar{g}_i$,
$\overline{N}_{ij}$, and $\overline{m}_i$ are physical scale independent
quantities to ${\cal O}(\alpha)$. $\bar{m}_i(\bar{m}_i)$ is the pole
mass of neutralinos. We also checked numerically the expression  
Eq.~(2) reproduce the full result very well.

\section{Numerical results}
\subsection{$m_{\tilde{Q}}$ Dependence of Various Cross Sections}
We next show the numerical dependence of the one-loop corrected 
cross-sections of $e^-e^+\rightarrow\tilde\ell_i\tilde\ell_j^*$
($\tilde\ell_i=(\tilde e_L^-,\tilde e_R^-,\tilde\nu_e)$) on the squark
mass.  We consider the case where the initial electron is completely
longitudinally polarized.  We therefore treat the following eight
modes,
\begin{eqnarray}
e^-_Le^+& \rightarrow& 
\tilde{e}_L^-\tilde{e}_L^+\ , \;
\tilde{e}_R^-\tilde{e}_R^+\ , \;
\tilde{e}_L^-\tilde{e}_R^+\ , \;
\tilde{\nu}_e\tilde{\nu}_e^*\ , \nonumber \\
e^-_Re^+& \rightarrow& 
\tilde{e}_L^-\tilde{e}_L^+\ , \;
\tilde{e}_R^-\tilde{e}_R^+\ , \;
\tilde{e}_R^-\tilde{e}_L^+\ , \;
\tilde{\nu}_e\tilde{\nu}_e^*\ .  \label{eq10}
\end{eqnarray}
The production involves the t channel exchange 
of chargino and neutralino, which depends of 
gaugino mass parameter $M_1, M_2$, higgsino mass parameter 
$\mu$, and $\tan\beta$.  
We take the three pole masses $(m_{\tilde{\chi}^0_1}$,
$m_{\tilde{\chi}^+_1}$, $m_{\tilde{\chi}^0_3}$), and $\tan\beta(M_Z)$
as inputs.  We assume $|\mu|\gg M_Z$, in which case $M^{\rm
eff}_1\simeq m_{\tilde{\chi}^0_1}$, $M^{\rm eff}_2\simeq
m_{\tilde{\chi}^+_1}$, and $|\mu^{\rm eff }|\simeq
m_{\tilde{\chi}^0_3}$ hold, where $M^{\rm eff}_1$, $M^{\rm eff}_2$,
and $-\mu^{\rm eff}$ are the (1,1), (2,2), and (3,4) elements of the
effective neutralino mass matrix $\overline{Y}_{ij}$, defined in
\cite{NPY}.

\begin{figure}[htb]
\centerline{
\psfig{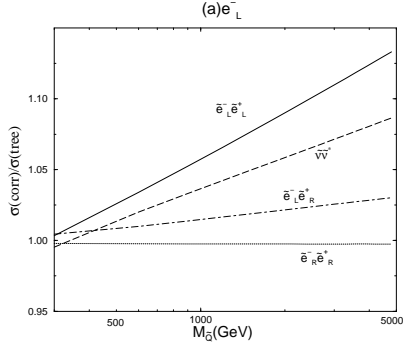}
}
\caption[fig3]{
The $M_{\tilde{Q}}$ dependence of the slepton production
cross-sections. Input parameters are 
$m_{\tilde{\chi}^0_1}=100$~GeV, $m_{\tilde{\chi}^+_1}=200$~GeV, 
$m_{\tilde{\chi}^0_3}=300$~GeV, $\tan\beta(M_Z)=4$, 
$m_{\tilde{\ell}}=200$~GeV, $A=0$, $\mu<0$, and $\protect
\sqrt{s}=500$~GeV. 
The corrected cross-sections are normalized by the tree-level
cross-sections\cite{NPY}.
}
\end{figure}

We show in Fig.~3 the $M_{\tilde Q}$ dependence of the cross-sections
for left handed electron beam.
Here we normalize the cross-sections to the tree-level values 
defined in Ref.\cite{NPY}. The one-loop corrected cross-sections 
of the modes which have a
$t$-channel contribution are similar to tree-level ones at
$M_{\tilde{Q}}\lsim 300$~GeV, but increase linearly with $\log
M_{\tilde{Q}}$.  Because the effective masses are equal to the input
pole masses, the
squark mass dependence of the one-loop corrected cross-sections is
primarily due to the difference between the effective theory gauge and
gaugino couplings. See Eq.(2)

The two channels which have $\tilde{W}$ contributions
($e^-_Le^+\rightarrow\tilde{e}_L^-\tilde{e}_L^+
,\tilde{\nu}_e\tilde{\nu}_e^*$) show the largest $M_{\tilde{Q}}$
dependence.  The  channel $e^-_Le^+\rightarrow\tilde{e}_L^-\tilde{e}_R^+$,
show smaller $M_{\tilde{Q}}$ dependence
from $\tilde{B}$ contributions. Nevertheless, 
we found these $M_{\tilde{Q}}$
dependences are significantly larger than the renormalization scale
dependence of the corrected cross-sections (see Fig.~2). In contrast, the
remaining channels, which have only $s$-channel contributions, show very
little $M_{\tilde{Q}}$ dependence, as explained in Section 2.

\subsection{Determination of $M_{\tilde{Q}}$}
The sfermion and chargino/neutralino
production cross-sections depend on  $M_1$, $M_2$, $\mu$, 
$\tan\beta$, and the sfermion mass
$m_{\tilde\ell}$.  The cross-sections also depend on the effective
fermion-sfermion-gaugino coupling $\bar{g}_{f\tilde{f}\tilde\chi}$,
which is nicely parameterized by $\log M_{\tilde Q}$.  By constraining
$\bar{g}_{f \tilde{f}\tilde\chi}$ we can determine $M_{\tilde Q}$, if
the rest of the parameters are known accurately enough.

It has been demonstrated that an accurate determination of
$m_{\tilde{\chi}}$ and $m_{\tilde\ell}$ is indeed possible if
sfermions $\tilde\ell$ are produced and dominantly decay into a
charged lepton $\ell'$ and a chargino or 
neutralino $\tilde\chi$\cite{JLC1, TSUKA}.
The
measurement of the end point energies determine the masses.
Recently, Baer et al. \cite{BMT} performed a MC
study for the case that left-handed sfermions are produced and decay
into a gaugino-like chargino or neutralinos. In their example called
point 3, $\tilde{\nu}_e\tilde{\nu}_e^*$ production is followed by
$\tilde{\nu}_e^{(*)}\rightarrow e^\mp\tilde{\chi}^{\pm}_1$.  The decay
mode $\tilde{\nu}_e\tilde{\nu_e}^*$ $\rightarrow
e^-e^+\tilde{\chi}_1^+\tilde{\chi}^-_1$ $\rightarrow e^-e^+ \mu 2j$
($\nu_{\mu}$ $2\tilde{\chi}^0_1)$ is background free and the
measured electron endpoint energies allow for a 1\% measurement of
$m_{\tilde{\chi}^{\pm}_1}$ and $m_{\tilde\nu}$.

The results of Ref. \cite{BMT} encourage us to consider their example
point 3. The chosen parameter set corresponds to $m_{\tilde{\nu}_e}=
207$ GeV, $m_{\tilde{\chi}^+_1}=96$ GeV, $m_{\tilde{\chi}^0_1}=45$~GeV
and $m_{\tilde{\chi}^0_3}=270$~GeV, and the lightest chargino and
neutralinos are gaugino-like. Their study suggests that we can take
$m_{\tilde{\chi}^+_1},$ $m_{\tilde{\chi}^0_1}$, and
$m_{\tilde{\nu}_e}$ as well constrained input parameters. For 20
fb$^{-1}$ of luminosity, their MC simulations show that at 68\% CL,
$(\delta m_{\tilde{\chi}^+_1},\ \delta m_{\tilde{\nu}_e}$) = (1.5~GeV,
2.5~GeV).

In the following we estimate the statistical significance of the
radiative correction to the production cross-section. 
We focus solely on the $\tilde{\nu}_e$ production
cross-section, because it is larger than 1 pb for a left-handed
electron beam, and larger than the other sparticle production
cross-sections at $\sqrt{s}=500$~GeV.

We would first like to provide a feel for the sensitivity to the
squark mass scale and $\tan\beta$ in the ideal case where we ignore
the slepton and gaugino mass uncertainties.  In Fig.~4(left) we show the
statistical significance of the loop correction by plotting
contours of constant cross-section. Here we fix the sneutrino mass and
determine $\mu,\ M_1$ and $M_2$ by fixing the one-loop corrected
masses $m_{\tilde{\chi}^0_1}$, $m_{\tilde{\chi}^0_3}$, and
$m_{\tilde{\chi}^+_1}$.  
We plot the contours
corresponding to the number of standard deviations of the
fluctuation of the accepted number of events.  The 1-$\sigma$
fluctuation corresponds to $\sqrt{N_{\rm input}}$, where $N_{\rm
input}$ is our nominal value of the number of events at
$M_{\tilde{Q}}=1000$~GeV and $\tan\beta(M_Z)=4$. The accepted number
of events $N$ is given by
\begin{equation}
N=A\cdot\sigma(e^-_Le^+\rightarrow \tilde{\nu}_e\tilde{\nu}_e)\times  
\left( {\rm BR} (\tilde{\nu}_e\rightarrow e\tilde{\chi}^+_1)
\right)^2 \times 100\ {\rm fb}^{-1}. 
\end{equation}
Here we took ${\rm BR}(\tilde{\nu}_e\rightarrow e \tilde{\chi}^+_1) =
0.6$ and overall acceptance $A=0.28$ .  The number of accepted 
events at our nominal point $N_{\rm input}$ is about 12800 for $\mu<0$.

\begin{figure}[htb]
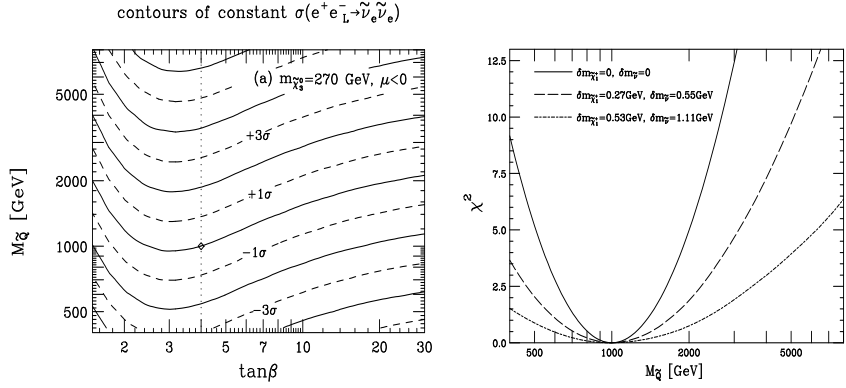

\centerline{
\psfig{file=fig4a.epsf,angle=90,height=144pt,width=160pt}
\hskip 0.3cm
\psfig{file=fig4b.epsf,angle=90,height=126pt,width=144pt}
}
\caption[fig4]{(left):The constraint on 
$M_{\tilde{Q}}$ and $\tan\beta$ coming from
$\sigma(e^-e^+\rightarrow\tilde{\nu}_e\tilde{\nu}_e^*\rightarrow
e^-e^+\tilde{\chi}^+_1\tilde{\chi}^-_1)$, with $\int dt L=100~{\rm
fb}^{-1}$. The central value is taken as $M_{\tilde{Q}}$=1000 GeV and
$\tan\beta(M_Z) =4$.$m_{\tilde\chi^0_3}=270$ GeV, $\mu<0$
(right):$\Delta\chi^2_{\rm min}$ vs. $M_{\tilde Q}$ with $\mu<0$ and
fixed $\tan\beta$, $m_{\tilde{\chi}^0_1}$ and $m_{\tilde{\chi}^0_3}$,
but allowing $m_{\tilde{\chi}^+_1}$ and $m_{\tilde{\nu}_e}$ to vary
freely.}
\end{figure}

We show the contours for $\mu<0$ in Fig.~4(left).  If $\tan\beta$ is well
measured, $M_{\tilde Q}$ is constrained to
$M_{\tilde{Q}}=1000^{+370}_{-280}$~GeV at 1-$\sigma$ significance.  If
instead we assume the constraint $2<\tan\beta<8$, the mild $\tan\beta$
dependence yields $700 < M_{\tilde{Q}} < 1900$ GeV. 
In the case $\mu>0$, the mixing of chargino 
gives rise to significant $\tan\beta$ dependence. 
Increasing the squark mass can be compensated for
by decreasing $\tan\beta$, and measuring sneutrino production then
determines a region of the $(M_{\tilde Q}, \ \tan\beta)$
plane. 

We now turn to the effect of the mass uncertainties. The sneutrino
production cross-section depends on the masses $m_{\tilde{\nu}_e}$,
$m_{\tilde{\chi}^0_1}$, $m_{\tilde{\chi}^0_3}$ and
$m_{\tilde{\chi}^+_1}$.    Of
these masses the cross-section is most sensitive to the sneutrino
mass.  All of the same chirality scalar production cross-sections
suffer from the strong $\beta_{\tilde\ell}^3$ kinematic dependence. 
Near threshold this results in an especially large sensitivity to the
final-state mass.  
Although a simple statistical scale-up of the results of 
Ref.~\cite{BMT} implies
a sneutrino mass uncertainty of only 0.3\%, this nevertheless leads to
a significant degradation in our ability to constrain the squark mass
scale. (Systematic errors might be the limiting factor here.)

Note, however, that the measurement of the sneutrino mass in
Ref.~\cite{BMT} was obtained by studying a small fraction of the total
sneutrino decay modes.  The mode they studied amounts to 
only about 4\% of the total
sneutrino decays. Using other modes, such as $e^-e^+4j
(2\tilde{\chi}^0_1)$, might reduce the mass error even
further. Because the $\tilde{\nu}_e$ production cross-section is
significantly larger than the other slepton cross-sections, isolating
the various sneutrino signatures is less affected by SUSY backgrounds
such as $e^-e^+\rightarrow \tilde{e}^-_L\tilde{e}^+_L$ $\rightarrow
e^-e^+ 4j(2\tilde{\chi}^0_1)$.

Now we show the constraint on the squark mass $M_{\tilde{Q}}$ after
taking into account the uncertainty of the masses $\delta
m_{\tilde{\nu}_e}$ and $\delta m_{\tilde{\chi}^+_1}$ for $\mu<0$ case. 
The effect of $\delta m_{\tilde{\chi}^0_3}$
turns out to be negligible for the case, and we assume it is possible to 
distinguish sign of $\mu$ by measuring heavier ino mass differences
at $\sqrt{s}>2m_{\tilde{\chi}^0_3}$.\cite{NPY}.
In Fig.~4(right) we plot $\Delta\chi^2_{\rm
min}$ against $M_{\tilde{Q}}$, where $\Delta\chi^2_{\rm min}$ is a
minimum of $\Delta\chi^2$ with respect to variations in
$m_{\tilde{\chi}^+_1}$ and $m_{\tilde{\nu}_e}$. The
region of $M_{\tilde Q}$ where $\sqrt{\Delta\chi^2_{\rm min}}<1,2,...$
corresponds to $1,2,...$-$\sigma$ error of the squark mass when the
chargino and sneutrino mass uncertainties are taken into account.  The
sneutrino mass uncertainty reduces the sensitivity of the production
cross-section to $M_{\tilde{Q}}$ considerably, because the effect of
increasing $M_{\tilde{Q}}$ can be compensated for by a small
increase in $m_{\tilde{\nu}_e}$.  On the other hand, we do not find
any significant effect due to non-zero $\delta m_{\tilde{\chi}^+_1}$.

{}From Fig.~4(right) we see that in this case, even with the sneutrino mass
uncertainty, we can reasonably constrain the squark mass scale. For
example, at the 1-$\sigma$ level with $M_{\tilde Q}=1$ TeV, we
constrain $M_{\tilde Q}$ to $1^{+1.2}_{-0.5}$ TeV, using the naive
scale up (from 20 fb$^{-1}$ to 100 fb$^{-1}$) of the statistical
errors of Ref.~\cite{BMT}.  This corresponds to the difference between
the gauge and gaugino effective couplings, $\delta g_2/g_2 =
0.011\pm0.006$. This can be compared to the estimate of the constraint
$\delta g_2/g_2 = \pm0.02$ from the chargino production measurement
\cite{CFPB}. Such comparisons are sensitive to different choices of
parameter space and other assumptions.   If
we reduce the mass uncertainties by a factor of 2, we find the
interesting constraint $600 <M_{\tilde{Q}}< 1500$ GeV.
\section{conclusions}

Supersymmetry is a beautiful symmetry which relates bosons and
fermions. If we wish to determine whether this symmetry is realized in
nature, the relations imposed between particles and their
superpartners must be confirmed by experiment.  Of course, discovering
a particle with the quantum numbers of a superpartner is the first
very important step in this procedure. An equally important test,
though, is the confirmation of the hard relations imposed by
supersymmetry, for example, the equivalence of the gauge and gaugino
couplings.

It has been argued that a next generation linear collider would be an
excellent tool to verify supersymmetry in this respect.  Production
cross-sections such as $\sigma(e^-e^+\rightarrow
\tilde{\ell}\tilde{\ell}^*)$ and $\sigma(e^-e^+\rightarrow
\tilde{\chi}^-_i\tilde{\chi}^+_i)$ involve the $t$-channel exchange of
gauginos or sleptons, so they depend on gaugino couplings. \cite{NFT,FPMT}

In this paper, we approached this problem from a somewhat different
direction. Because supersymmetry must be badly broken by soft breaking
terms, the tree-level relations of the couplings are also broken, by
radiative corrections. The corrections are logarithmically sensitive
to the splitting of the supersymmetry multiplets. To quantify this, we
have calculated the full one-loop correction due to (s)quark loops of
the slepton production cross-sections.  The difference between the effective
lepton-slepton-gaugino couplings
$g^{\rm eff}_{\tilde{G}_i}$ and the effective gauge
couplings $g_i^{\rm eff}$ is given by a coupling factor times
$\log M_{\tilde{Q}}/m_{\tilde\ell}$.

We gave an explicit example which illustrates that the statistics at
the future linear collider may be enough to constrain the squark mass
scale through the measurement of the slepton production cross-section.
We found, with 1-$\sigma$ significance, $M_{\tilde Q}$ could be
constrained to $1^{+1.2}_{-0.5}$ TeV by the measurement of the
sneutrino production cross-section.   We found this constraint in the
$\mu<0$ case where we took into account the errors (based on existing
MC simulation) of the sneutrino and light chargino masses, but assumed
$\tan\beta$ was well constrained by other measurements.

The mass of the sleptons and gauginos must
be measured very precisely in order to successfully constrain the
squark mass scale via production cross-section measurements. 
In order to determine the ultimate
sensitivity of this procedure, a thorough study of the systematic
uncertainties in the slepton mass measurements is necessary. 

It is important to note that the constraint on the squark mass scale
can be stronger than the one presented in this paper.  Here, we
estimated the sensitively to squark mass scale by utilizing sneutrino
production followed by its decay into a chargino and an electron.
Depending on the spectrum and center-of-mass energy, there will
typically be many other production processes which involve $t$-channel
exchange of gauginos or sleptons, and all those amplitudes have
$\log M_{\tilde{Q}}$ corrections.

The constraint on the squark mass we have realized here could be
unique in the sense that this information may not be available at the
LHC. Even if the LHC squark production rate is large, the gluino could
be produced in even larger numbers, creating a large irreducible
background to the squark signal. A large gluino background
could make the extraction of the squark mass from kinematical
variables difficult.

On the other hand, if information on the squark masses is obtained at
the LHC, we would have rather accurate predictions for the gaugino
couplings.  In this case, the measurement of the production
cross-sections we considered here would constrain new
supersymmetry-breaking physics with standard model gauge quantum
numbers.  In a sense, the study proposed here is similar in nature to
studies performed at LEP and SLC. The physics of gauge boson two-point
functions has been studied extensively at LEP and SLC, and it has
provided strong constraints on new physics.  Similarly, a future LC
and the LHC might provide precision studies of the gaugino two-point
functions, to realize a supersymmetric version of new precision tests.

\end{document}